\documentclass{article}

\usepackage{arxiv}

\usepackage[utf8]{inputenc} 
\usepackage[T1]{fontenc}    
\usepackage{hyperref}       
\usepackage{url}            
\usepackage{booktabs}       
\usepackage{amsfonts}       
\usepackage{nicefrac}       
\usepackage{microtype}      
\usepackage{lipsum}
\usepackage{graphicx}
\usepackage{amsmath}
\usepackage{multirow}
\graphicspath{ {./images/} }

\title{Agent-Based Simulation of UAV Battery Recharging for IoT Applications: Precision Agriculture, Disaster Recovery, and Dengue Vector Control}

\author{
 Leonardo Grando (*) \\
  School of Technology\\
  State University of Campinas, \\
  Limeira, São Paulo, Brazil 13083-872 \\
  (*) Corresponding author\\
  \texttt{lgrando@gmail.com} \\
   \And
    Juan F Galindo Jaramillo \\
  School of Technology\\
  State University of Campinas, \\
  \texttt{jgalindoj@ieee.org} \\
  \And
 Jose Roberto Emiliano Leite \\
  School of Technology\\
  State University of Campinas, \\
  \texttt{joserobertoemilianoleite@gmail.com} \\
  \And
  Edson Luiz Ursini \\
  School of Technology\\
  State University of Campinas, \\
  \texttt{ursini@ft.unicamp.br} \\
}

\begin{document}
\maketitle
\begin{abstract}
The low battery autonomy of Unnamed Aerial Vehicles (UAVs or drones) can make smart farming (precision agriculture), disaster recovery, and the fighting against dengue vector applications difficult. This article considers two approaches, first enumerating the characteristics observed in these three IoT application types and then modeling an UAV’s battery recharge coordination using the Agent-Based Simulation (ABS) approach. In this way, we propose that each drone inside the swarm does not communicate concerning this recharge coordination decision, reducing energy usage and permitting remote usage. A total of 6000 simulations were run to evaluate how two proposed policies, the BaseLine (BL) and ChargerThershold (CT) coordination recharging policy, behave in 30 situations regarding how each simulation sets conclude the simulation runs and how much time they work until recharging results. CT policy shows more reliable results in extreme system usage.  This work conclusion presents the potential of these three IoT applications to achieve their perpetual service without communication between drones and ground stations. This work can be a baseline for future policies and simulation parameter enhancements.   
\end{abstract}


\section{Introduction}
The Internet of Things (IoT) enables the automation of several sectors of economics, including Precision Agriculture, disaster recovery, and mosquito (Dengue) disease combat/detection. Initially, this work focused on three types of drone application descriptions, and the main issue was the relatively short duration of UAV batteries.

This work focuses on two approaches: i) Necessary characteristics of the three applications for IoT UAVs, and ii) deploying a drone's battery recharging coordination process using Agent-Based Modeling and Simulation (ABMS). This approach considers that drones do not communicate between them about their recharging decisions, as in El Farol Bar Problem \cite{Arthur1994}.

This article aim to propose a process of temporal coordination of the recharging process of a swarm of drones to increase its autonomy. These drones may perform tasks in precision agriculture activities, disaster recovery, and fighting against dengue vectors. To this end, we proposed a strategy development to coordinate these drones in the decision to recharge or continue their work. This work approach uses a Game Theory approach called the El Farol Bar Problem and will be compared to a baseline strategy. This strategy considers the standard behavior of recharging when the amount of energy reaches a minimum limit, such as a warning from a cell phone when its battery reaches a minimum value. 

The authors don't know a real-world autonomous swarm with several drones in large numbers as our simulation environment setup considers 100 UAVs in swarms, so we can't compare our model with real-world applications. For this, we seek to model the environment and evaluate a sensitivity analysis to verify and validate the parameters and variables developed in the model. We experiment extensively (60 simulation sets) and replicate 100 times each to ensure statistically reliable results.

Two results were present in this work. The description of agriculture, disaster recovery, and Dengue disease combat application using drones. The second result is a drone energy supply procedure modeling an Agent-Based simulation description. The simulation considers two approaches in the recharging coordination and the reliability and efficiency of extensive simulation (6000 simulation runs).

This article structure is: Section 2 presents the characteristics and trends of drone technology and agent-based simulation. Section 3 addresses the IoT Applications that are the main focus of this work: smart farming, disaster recovery, and fighting dengue, showing its current needs, developments, and trends. Section 4 introduces the agent-based modeling and simulation of recharging drone batteries and details of the Work Development: location, variables, code, and environment analysis. Section 5 describes the results and discussions about the values obtained. Finally, Section 6 presents conclusions and considerations, research limitations, and suggestions for future work.

\section{Related Works}
\label{sec:headings}

This chapter presents selected literature that presents drone technology, its classification, and power supply modes.

\subsection{Drones Technologies}

Unmanned aerial vehicles (UAVs) are aerial vehicles that are not piloted by humans. They can be remote-controlled to perform their tasks, carry out pre-programmed missions, or carry out actions autonomously \cite{Fahlstrom2012}.

\begin{figure}[ht]
	\centering
	\includegraphics[width=0.5\textwidth]{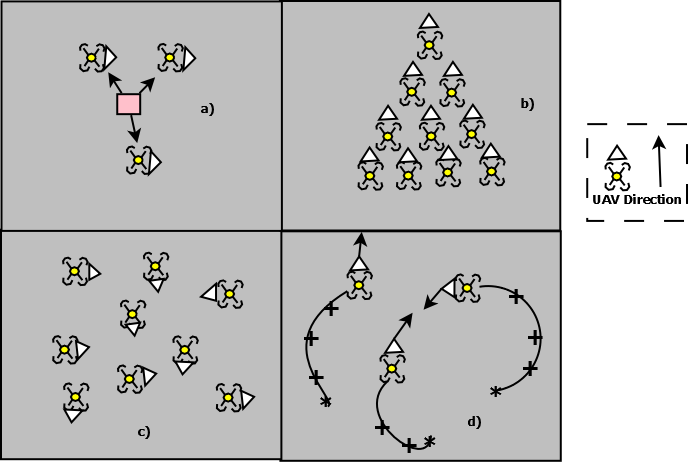} \\
    \caption{Graphical representation of UAV architectures. Based on: \cite{Fahlstrom2012}}
	\label{fig:1}
\end{figure}

Figure \ref{fig:1} illustrates a generic UAV system with its Ground Control Station, the data link antenna, the UAV, and a satellite \cite{Fahlstrom2012}. Drone flight duration can be increased by improving the technology used in the battery or supplying energy to these batteries (via recharging or battery exchange).

Drones can act alone or together (multi-UAV), with the advantages of using a swarm of drones instead of a single drone \cite{Valavanis2015}: 
\begin{itemize}
    \item Possibility of using the swarm to carry out multiple and simultaneous interventions;
    \item Greater efficiency in the action’s execution time;
    \item Obtain a more reliable system due to redundancy;
    \item The cost of using small devices can be lower than just big ones, as the weight and size of drones increase energy consumption.
\end{itemize}

\begin{figure}[ht]
	\centering
	\includegraphics[width=0.5\textwidth]{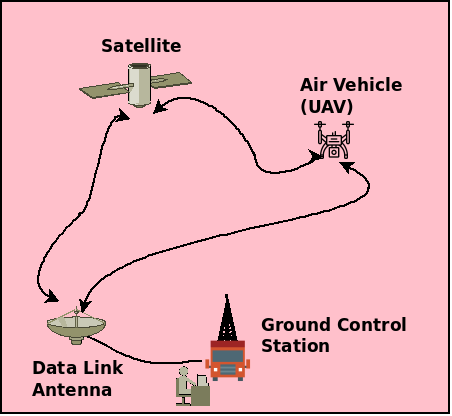} \\
    \caption{Four types of Multi-UAV architectures. Based on: \cite{Valavanis2015}}
	\label{fig:2}
\end{figure}

When considering a drone swarm, their architectures can be classified into four types \cite{Valavanis2015}.:
Physical coupling: When UAVs are connected with physical links and their movements are limited by forces that depend on other types of UAVs - Application example: transportation of a single object by multiple autonomous vehicles. (Figure \ref{fig:2}-a);
Formation: The UAVs are not physically connected, but their movement is restricted to maintain the predefined formation (Figure \ref{fig:2}-b).
Swarms: Homogeneous vehicles team whose interaction generates collective emergent behavior. They have decentralized control. (Figure \ref{fig:2}-c) 
Intentional Cooperation: The team of UAVs moves according to trajectories defined by individual tasks allocated to execute a global mission. (Figure \ref{fig:2}-d).

The key concept of swarm behavior is that global complex collective behaviors emerge through the simple interaction between large relative numbers of unintelligent agents. Agent-based simulation models are widely used to simulate complex systems \cite{Wilensky2015, Mitchell2009}. 

Consider a swarm of drones, in which each drone should decide whether it should carry out its work or recharge its batteries in each simulation run. The idea of a swarm is that there is no communication between drones to minimize battery consumption, at least in most missions (as occurs with people in El Farol Bar Problem). In critical cases, for example, when the route could cause a collision, it is foreseen that there must be communication between the drones.

\subsection{Drone classification}

Drones can be classified by their maximum takeoff weight (MTOW), degree of autonomy, and operational altitude \cite{Valavanis2015}. The Brazilian MTOW classification \cite{ANAC2023a} considers three MTOW classes. Class 1 was UAVs with more than 150 kg; Class 2 considers UAVs with MTOW between 25 and 150 Kg; and Class 3 corresponds to UAVs less than 25 Kg. The drone sizes can be classified into Very Small UAVs, Small UAVs, Medium UAVs, and Large UAVs. The range capacity classification considers categories as Very Low-Cost Close Range, Close Range, Short Range, Mid-Range, and endurance \cite{Fahlstrom2012}.
Another UAV classification considers the number of rotors. In agriculture applications, drones with less than three rotors are not applied due to aerodynamic and control factors. Drones with four rotors were used more. Drones with six or eight rotors were also applied, mainly in applications that carry a larger payload. These drones have the disadvantage of being more difficult to control. Regarding their autonomy, battery-powered drones can fly for an average of about 20 to 30 minutes \cite{Mohsan2022a}. Drones can be attached with several sensors and payloads for their movement and communication \cite{Rahman2021} as Table \ref{tab:1}.

\begin{table}
    \centering
        \caption{UAV payload elements and applications, based on \cite{Rahman2021}}
    \begin{tabular}{cc}
    \hline
         Name of the Element & Objective \\
         \hline
         Camera (RGB/Infrared/Thermal) & Capturing Images\\
         GPS & Navigation \\
         WSN (Wireless Sensor Network) & Monitoring Circumstances\\
         Altimeter & Measuring Altitude\\
         Accelerometer & Acceleration Grading\\
         Gyroscopes & Maintaining Orientation and Angular Velocity \\
         Magnetometer & Measuring Magnetic Field Strength and Direction\\
         Battery & Retaining Power\\
         BLDC (Brushless DC Electric Motor) & Movement Control \\
         \hline
    \end{tabular}
    
    \label{tab:1}
\end{table}

\subsection{Energy Supply}

\begin{figure}[ht]
	\centering
	\includegraphics[width=0.5\textwidth]{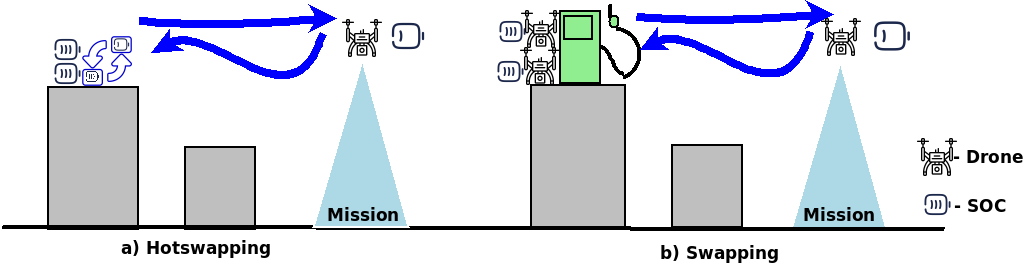} \\
    \caption{Battery exchange techniques, based in \cite{Campi2019a}.}
	\label{fig:3}
\end{figure}

Drones can use higher-capacity batteries, but this can increase their mass, negatively influencing UAV flight capacity\cite{Mohsan2022a}. As the drone battery capacity increases, its total mass usually increases as well. If the relationship between battery capacity and mass exceeds a limit, the battery life of the drones begins to reduce \cite{Jain2020}. The use of new materials may result in increased battery energy density. Other technologies like super capacitors7, photovoltaic power supply \cite{Aissi2020, Al-Obaidi2018, Avila2018}, fuel cells \cite{truog_insights_2020,Gong2018}, or hybrid forms of energy \cite{Avila2018}, have also been researched. The power supply can also be carried out via cable \cite{Fauzi2022} or by a wireless power transfer (WPT) system \cite{Campi2019a}.
In this work, we consider the internal quantity of charge of the Battery as Status of Charge (SOC) values. This range is between 0 to 100\%, if drones achieve a SOC of less than 0\%, they will stop work by starving. Figure \ref{fig:3} shows the battery replacement process. The hot-swapping process (Figure \ref{fig:3}-a) is when drained batteries are exchanged for charged batteries, allowing the drones to continue working. In the swapping process (Figure \ref{fig:3}-b), drones with drained batteries are replaced by other recharged drones and enter the recharging process. A swapping process is composed of three elements: i) the battery swap station, ii) the available batteries, and iii) a control system to manage the swarm of UAVs \cite{Mohsan2022a}.

Regarding the drones recharging coordination process, a Systematic Literature Review in agriculture and disaster context \cite{Grando2025} found three research gaps about knowledge, methodological, and practical areas. Because of practical applications in these areas motivate us to perform this simulation work. The \cite{grando_modeling_2024} work evaluates different simulation values for this work application.

\subsection{Simulation and Complex Systems}

A complex system is composed of multiple individual elements that interact with each other, and their behavior or properties are not predictable from each component. A complex system with broad networks of components with no central control and simple rules of operation gives rise to complex collective behavior (no leader), sophisticated (internal and external) information processing, and adaptation via learning or evolutionary processes \cite{Wilensky2015, Mitchell2009}.

A complex system has at least one of six characteristics. These characteristics are interactions between their components, the emergence, where the whole results are more than the sum of its parts, long-term behavior system dynamic results, self-organization of the whole system, adaptation and evolving capacity, and Interdisciplinary \cite{DeDomenico2019} 

Agent-based modeling (ABM) is a computational way of modeling complexity. That considers entities called agents, a place where these agents interacted, and their interactions. The ABM goal is to create agents and rules that will generate a target behavior. Three elements compose an agent-based model: the agents, the environment, and their interactions. \cite{Wilensky2015}

NetLogo \cite{Wilensky1999} is an Agent-Based Simulation Software that provides a simple/easy model development effort, with a medium/broad modeling strength and with active objects with simple goals implemented as mobile agents (turtles, patches, links, and the observer) \cite{Abar2017}.

\begin{figure}[ht]
	\centering
	\includegraphics[width=0.5\textwidth]{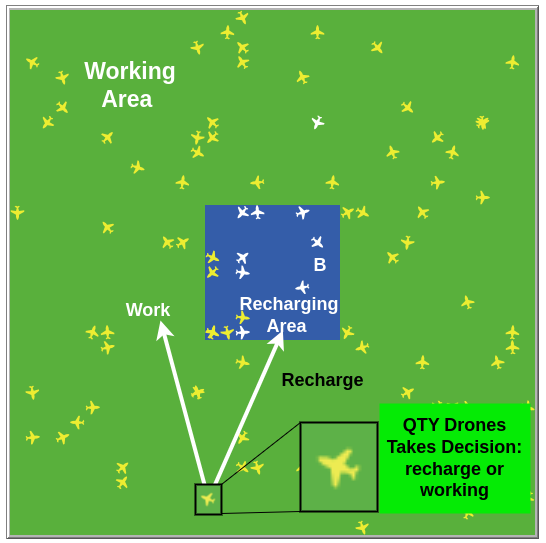} \\
    \caption{Agents’ decision process in each simulation cycle.}
	\label{fig:4}
\end{figure}

Figure \ref{fig:4}, shows how we used these elements in our model, where the agents (turtles) are the drones, the simulations patches that define the environment. The working and recharging patch areas and the agents interact with the environment when they recharge. 

\subsection{Recharging simulation in a swarm}

Concerning the modeling of drone swarms and recharges of our best knowledge, two works focus on coordinating a swarm of drones for its continuous operation. The authors \cite {BoggioDandry2018} consider a finite state analysis considering seven states. Each drone chooses whether to change after assessing its situation.  The authors \cite{Grando2020} used a model based on the El Farol bar for recharging drones. Our work follows a similar approach, adding more details as the amount of energy used for each drone and proposing new recharging decision-making policies. Using Agent-Based Simulation for modeling inductive behavior, as in the El Farol Bar model, it is possible to improve recharging decision policies. 
Therefore, our current work focuses on coordinating the power supply for drones, seeking to improve their capacity to complete their missions. This work considers that the Swarm systems can perform some level of their Autonomy Control Level (ACL), at least Level I, meaning executing a pre-planned mission capacity \cite{Dalamagkidis2015}. 

\section{IoT Applications Analyzed}

This work considers three types of applications based on drones: Smart Farming, Disaster Recovery, and the Fight Against Dengue Vectors that can use battery-recharging procedures.

\subsection{IoT Smart Farming Application (Precision Agriculture)}

\begin{figure}[ht]
	\centering
	\includegraphics[width=0.5\textwidth]{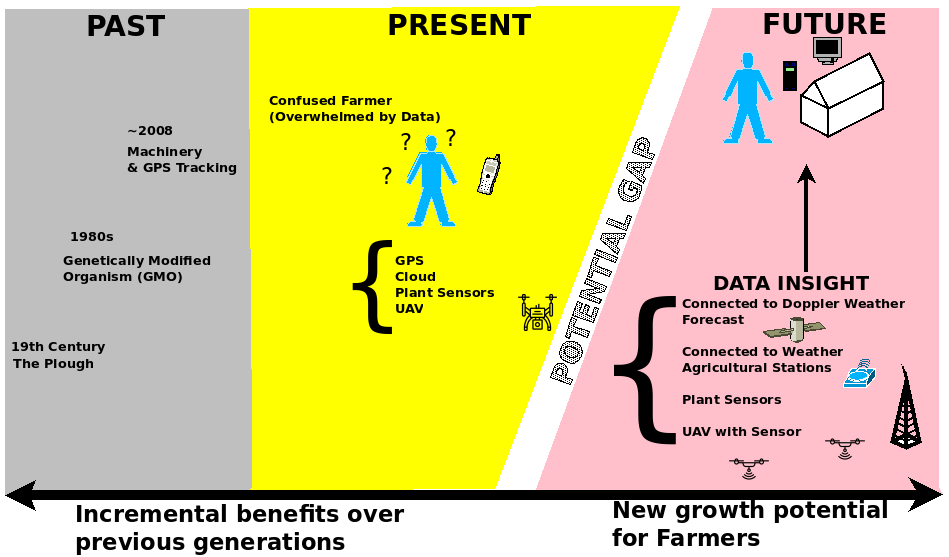} \\
    \caption{Past, present, and future of the agriculture technology, based on \cite{Fathallah2017}.}
	\label{fig:5}
\end{figure}

The agriculture sector impacts the Brazilian economy \cite{FAO2022}. The automation of agriculture activities may increase their productivity sustainably, reducing the planting area and avoiding deforestation (not ecologically acceptable). Drones are robotic tools that, in addition to other automated machines such as harvesters, irrigators, and planters that can become autonomous, may help to increase agricultural productivity. Drones allow farmers to seed and spray more homogeneously and precisely. The farmer will be able to identify the functioning of irrigation equipment, possible failures in crop coverage, and water stress. It can also help monitor animals and livestock equipment, locate animals, identify sick animals, read chips, and monitor fences and water fountains.  Figure \ref{fig:5} shows the precision agriculture evolution, based on \cite{Fathallah2017}. 
IoT can be used in several areas in Precision Agriculture (PA) as monitoring of soil, water, air, crops, and animals; the control of irrigation, fertilizers, pesticides, lighting, and access; in the prediction: of environmental conditions, estimated production and growth of the plantation and the Logistics: trade and transport.

\begin{figure}[ht]
	\centering
	\includegraphics[width=0.5\textwidth]{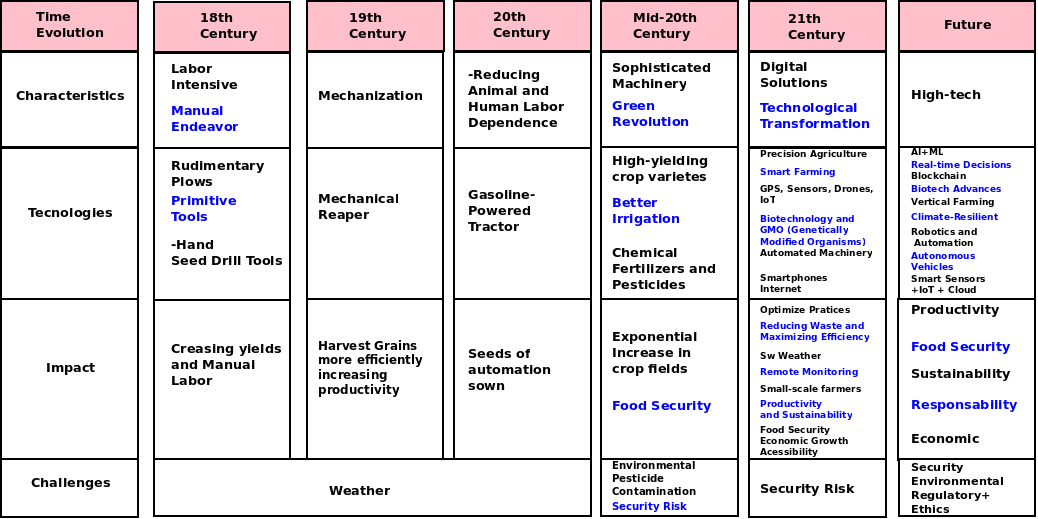} \\
    \caption{Evolution of precision agriculture.}
	\label{fig:6}
\end{figure}

Fathallah et al. \cite{Fathallah2017} present the evolution of agriculture in recent centuries, from the use of the plow with animal power through agricultural seed research and development to the present day with the use of cell phones, mobile networks, sensors, and IoT, as shown in Figure 5. This is how precision agriculture emerged. PA eases decision-making regarding factors such as the amount of water or fertilizers to use in each sector of the plantation. They also present a sensor network with an access card that gives access to data and shows which information is sent through the network to cloud applications to assist in better decision-making.
Agriculture involved intensive manual work and primitive tools for a long time, with the responsibility to feed the population. This area has undergone technological transformations in recent centuries, going from plows and hand tools of past centuries to today's automatic machines and digital solutions. Figure \ref{fig:5} shows the historical changes that began in the 18th century with the usage of seed drill tools. In the 19th century, the mechanical reaper appeared, facilitating harvesting and productivity increases. In the 20th century, gasoline-powered tractors appeared, reducing dependence on animal traction and human labor, the seeds of automation. In the middle of the 20th century, new technologies and practices appeared, such as high-yielding crop varieties, better irrigation systems, and fertilizer and pesticide use, increasing harvest productivity.
In the 21st Century, many technologies appeared in agriculture: GPS, Sensors, and Drones, transforming traditional agriculture into PA (Smart Farming) with IoT, as shown in Figure \ref{fig:6}. PA reduces the waste of water, fertilizer, and pesticides. Genetically Modified Organisms (GMOs) also appeared, generating seeds that were more resistant to pests. Automatic Machines also appeared (harvesters, automated irrigation systems, and self-driving tractors). 
Information and Communication Technologies grant plantation/soil data collection and remote monitoring by the climate data and prediction algorithms. The Internet allowed remote access to cloud databases. All these technologies led to an increase in productivity, reducing waste and improving environmental sustainability, population food security, economic growth, and small producer competitiveness. Technologies such as Artificial Intelligence, Machine Learning, Blockchain, Biotechnology Advances, Vertical Farming with planting layers with water reuse, climate-resilient agriculture, robots, and autonomous vehicles will drive the PA future.

\begin{figure}[ht]
	\centering
	\includegraphics[width=0.5\textwidth]{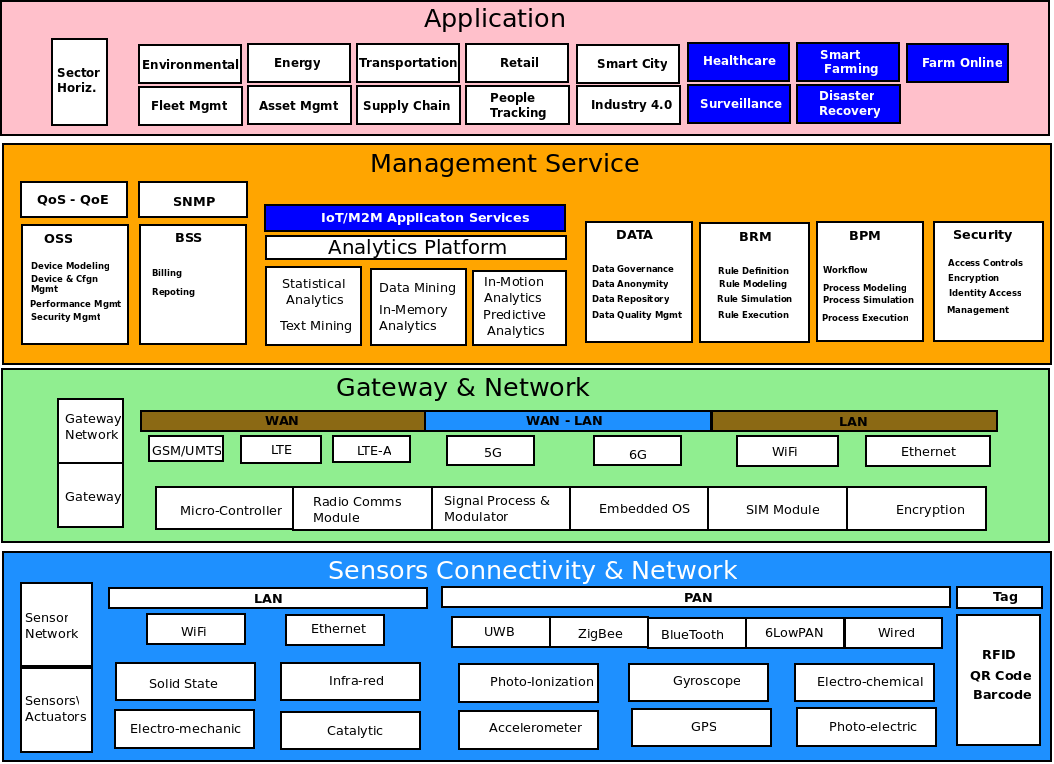} \\
    \caption{IoT Architecture based in \cite{Patel2016}.}
	\label{fig:7}
\end{figure}

Figure \ref{fig:7} shows the Layer stratification of an IoT architecture \cite{Patel2016} used in agriculture. The Application Layer is responsible for IoT applications, such as Smart Farming, Smart City, Smart Grid (Electrical Energy), Disaster Recovery, etc. The Service Support and Application Support Layer is responsible for Service and Application Support. The Network Layer is responsible for network communication and transport of information to the destination: Internet Networks, 4G/5G Mobile Networks, RFID Networks, WSN Sensor Networks (AD HOC), Zigbee Networks, and Optical Networks. The Device Layer is responsible for sensing information generation: Sensors, Sensor Networks, RFID Tags, GATEWAYS, Drones, GPS, Actuators (sirens, alarms, panels with escape routes, smartphones), and Spray Planes.
Sensors and meter usage allowed IoT tools to improve their capabilities in sensing (monitoring), decision-making with the collected data, and triggering actions for the actuators. Evolution in this process has been occurring in recent years through the use of GPS, drones, optical fiber (as sensors), and smartphones (as warning actuators and access to data).
Ad hoc and Zigbee Sensor Networks, RFID (identification of badges and equipment), sensors, meters, and actuators can analyze stored data for decision-making. The data collected by sensors, meters, and RFID are sent through telecommunications networks (Ad hoc, RFID, Internet) to the database with cloud applications, where the data is analyzed with predictors to detect any failure.

In agriculture, drones can obtain information through images or by collecting data using sensors. This information can help make data-driven operational decisions. Works \cite{Fathallah2017, Hartanto2019, RadoglouGrammatikis2020} describe drone applications for PA as herds and crop monitoring, image capturing, seeding, fruit harvesting, and spraying. Brazilian Law \cite{2021} regulates the UAV operation for spraying, spreading, and fertilizing in Brazil.

Rahman et al. \cite{Rahman2021} present different uses of drones in agriculture and some limitations as privacy risks, complex spraying environments, and long-distance positioning. Unexpected users can compromise the data link connection between drones and the base station. Another recommendation is not to use nano and micro drones because of their lower battery and weight-load capacity. This difficulty in guiding drones remotely may be another factor favorable to the use of a swarm and a decentralized recharging approach, as our work proposes.

\begin{figure}[ht]
	\centering
	\includegraphics[width=0.5\textwidth]{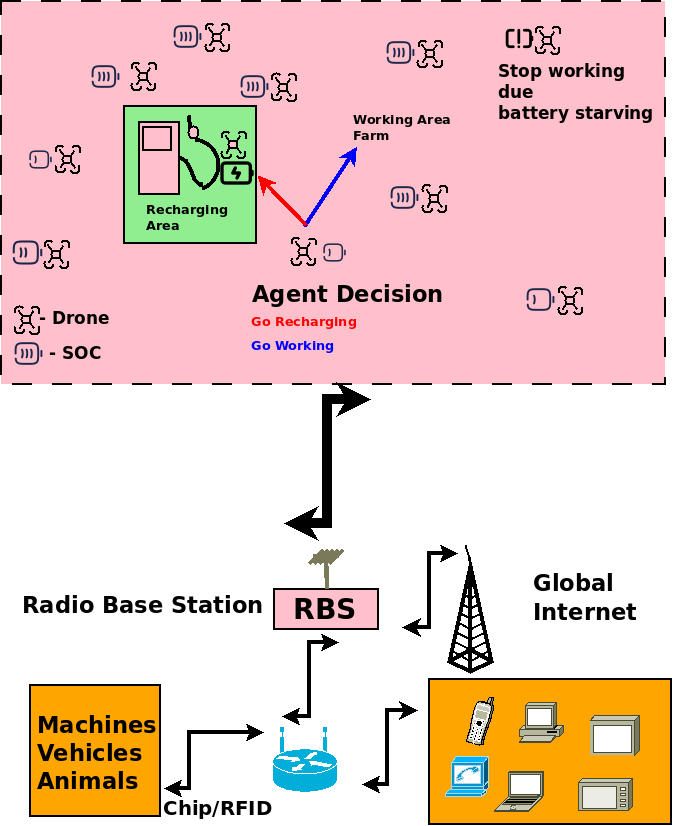} \\
    \caption{IoT Architecture based in \cite{Patel2016}.}
	\label{fig:8}
\end{figure}

Drones can image, harvest fruit, fertilize, water, and spray insecticide with precision in the correct quantity and location. We suggest a connected farm that links all this equipment to a Radio Base Station (RBS). This base connects the farm to the Internet. We present how this communication works in Figure \ref{fig:8}.
The 4G and 5G networks are essential to employees and equipment using the RBS, including communicating with sellers, buyers, and external cooperatives. The connected farm thus becomes an Industry 4.0 with seed-planting tractors, harvesters, irrigators, and autonomous drones. Also, RFID (Radio Frequency Identification) chips can be used in animal earrings and necklaces as employee and machine badges.

Every sensor can use an attached embedded system that adds intelligence with processor, memory, and radio communication, thus forming the Sensor Network. The information destination is for the cloud application to make the best decision for the collected data.

\subsection{IoT Disaster Recovery Application}

\begin{figure}[ht]
	\centering
	\includegraphics[width=0.5\textwidth]{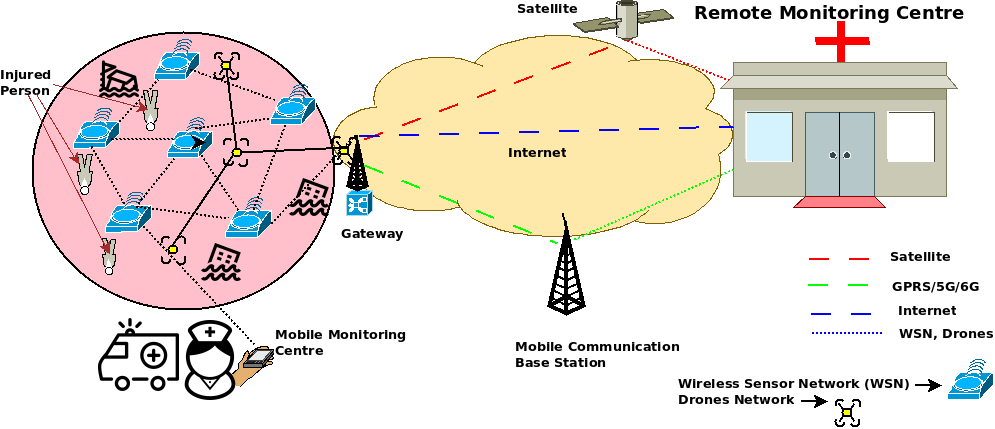} \\
    \caption{Disaster Recovery Scenario based in \cite{ITU2014}.}
	\label{fig:9}
\end{figure}

Drones and 4G and 5G networks are tools that help in natural disaster recovery processes. A recovery process can occur during and after the disaster event \cite{ITU2014}. Earthquakes, tsunamis, storms, dam failures, hillside falls, and flooding are typical and increasingly frequent cases due to global warming. Smartphones can warn the population of the occurrence of the disaster and possible escape routes. Drones film and show the areas most affected by the disaster, in addition to giving audible alarms to the population indicating the disaster and escape routes. Photos and footage made by drones can be used to compare the “Before And After” showing the spread and severity of the disaster. The information flow can be improved using smartphone messages before, during, and after the event. Modeling and simulating the recharging and duration of drone batteries also provide important observations for achieving the critical missions assigned to drones. Figure \ref{fig:9} shows a disaster recovery scenario using IoT.

Horio et al. \cite{Horio2019} present a proposal to simulate a disaster scene using drones. Drones can perform various activities such as fire and rescue detection, firefighting, injured rescue, and information transfer, and drones also recharge their batteries if necessary. That article uses the Bias and Raising Threshold (BRT) algorithm as a policy. In BRT, distributed agents can choose the best among options, without the presence of a leader. As in the case of the Arthur \cite{Arthur1994}, the agent's communication is limited. They have a scoring system to emulate the health and performance of each drone.

The use of environmental sensors (humidity, displacement, pressure, temperature) and vital signs sensors (smartwatch: temperature, blood pressure, heart rate, blood glucose level, and blood oxygen) to anticipate the discovery of an accident and a patient's need for care \cite{ITU2014}. Disasters with many deaths are mainly due to population alert failure. To alert and guide the population, smartphones, sirens, and alarms can act as actuators to warn the population and government bodies (government, state, city hall, fire department, and civil defense). Also, escape route panels, cell phone signals, and drones can help show the best evacuation path for disaster victims, reducing the number of fatalities. 
Drones can also identify the locations most affected by the disaster and the existing alternative paths to reach the population of that location, and warn the population to leave the risk area.

\subsection{Use of drones to combat dengue}

Regarding the fight against dengue, literature reports two main approaches to the use of drones: i) monitoring new breeding sites for the insect through pictures \cite{Amarasinghe2020, DeMesquita2021}, and ii) the use of drones to disperse sterile Aedes aegypti mosquitoes \cite {Ackerman2017, FAPESP2024}.
Brazil is on alert regarding the dengue virus epidemic caused by the proliferation of outbreaks of Aedes aegypti mosquitoes. Considering the first dengue symptom incidence, Brazil recorded 6264 deaths and 6,601,253 probable cases during 2024 \cite{Brasil2024}.
The explosion of dengue cases occurred early this year in Brazil due to climate change, increasing temperature, and rain. This insect is also the vector for other infectious tropical diseases, such as Chikungunya and Zika virus \cite{Amarasinghe2020}. About 75\% of mosquito breeding sites are found in homes \cite{Brasil2024b}, making it necessary to constantly monitor the land, backyards, gardens, and roofs of residences. Drones can make this pattern recognition about dengue spots.
Therefore, our simulation can also evaluate how drone swarms can improve their flight time by searching for more dengue spots and delivering more sterile Aedes aegypti mosquitoes.

\section{Development}

To evaluate the applications previously proposed, we created a study to abstract specific details of a drone swarm system using agent-based modeling, valuing the essential characteristics to ensure that the mission is carried out by studying conditions that allow drones to coordinate their recharging decision to optimize their flight range.

Drones (agents) can be considered as autonomous agents \cite{Macal2016}. Agents (drones) are heterogeneous, they make decisions autonomously, but they do not interact with each other.

\subsection{The El Farol Bar model}

The El Farol Bar problem, created by William Brian Arthur \cite{Arthur1994}, aims to simulate inductive reasoning (bounded rationality). This reasoning is as opposed to deductive and logical behavior as is traditional theoretical economics problems. 
El Farol is a bar in Santa Fe, USA, a neighborhood with a certain number of agents $(QTY)$. They want to decide whether to go to this bar. This decision is based on the previous occupancy value history. The occupation limit uses a happiness criterion called the overcrowding threshold $(B)$. When the bar occupancy is lower than $B$, bar attendees will enjoy a good night. This agent uses a set of k autoregressive predictors that use a window of previous values of the bar's occupancy in $m$ previous weeks to estimate the next agent's predicted attendance value. If the agent's predicted value is less than $B$, these agents will attend the bar in the next simulation round. 

All of these decisions don't collude with other agents (there is no collusion among them).  
Our model uses the El Farol NetLogo model \cite{Rand2007} library as a base for the simulation implementation. We use the same autoregressive model whose estimators' internal weights were defined at the beginning of the simulation. We also use this approach to calculate predicted values in our simulation model.

In our ABM, we consider the drones as agents, the charging station as the El Farol Bar, and the predictors as a source of calculations so that the agents' internal policies can make their decisions. El Farol Bar Problem is already used in resource congestion problems \cite{Bell1999, Sharif2011}.
At first, we focused on two problems to be minimized by this strategy: 
\begin{enumerate}
    \item the reduction in communication between drones (which can occur in other critical situations) with the consequent reduction in battery consumption;
    \item Mitigate the need for remote control to recharge drones during continuous work on farms or disasters in the recharging process.
\end{enumerate}

This simulation model was developed in the agent-based simulation software NetLogo \cite{Wilensky1999}. Our proposed model considers the coordination of the decision process without collusion between agents (they do not communicate) about their decision, being an advantage in adverse environments such as agriculture or disaster recovery and in reducing the process of consumption of computational and communication resources to make this decision.

The simulation objective is to develop a process for coordinating the decision to recharge a swarm of drones, where they will define whether or not to recharge their batteries. To make this decision, each agent has internal estimators, which we call decision policies. The complexity of the calculations performed by these estimators is varied, and their input data can range from the amount of energy remaining in each agent's battery, the baseline (BL), and our proposed  Charger Threshold (CT) Policy to the previous value of the number of drones that visited the charging station (CT).

\subsection{Variables and Decision Process}

In this model, a $QTY$ number of drones needs to decide which action should be taken when choosing between visiting the work environment or the charging station. The first location is where it performs its mission (spraying, obtaining images, serving as a source of connectivity in case of disasters), and the second place is its charging station. This charging station has a limited capacity, with this capacity being a fraction of the drone QTY values.
The variables in this simulation model can be related to the problem's demand, the system's recharge capacity (supply), and the agents' decision process.
Concerning the demand for the swarm of drones, they are related to the Quantity of Agents (Quantity - $QTY$), the average energy expenditure of each agent (Battery Consumption - $BC$), and the standard deviation of BC ($SD$).
Regarding the recharging offer for drones, the capacity of the recharging station $(B)$ and the amount of energy supplied with each recharge carried out (Battery Gain - $BG$) are the considered variables.
The model has two decision-making policies, the base policy ($BL$) and the Charger Threshold (CT) policy. Figure \ref{fig:8} presents graphically the decision process of both recharging processes. The BL policy considers the amount of energy ($SOC$) each agent has to define whether or not to recharge, analogous to the decision to fill up a car or recharge a cell phone when reaching a critical limit. This critical limit is called the Lower Reload Limit ($LW$).

CT Policy considers both the amount of energy present in the drone battery and the decision process based on the El Farol Bar, based on the history of agents who went to recharge at the charging station to make a decision. Regarding the history of agents that have tried to recharge their batteries in the last m weeks, this information was sent by the Radio Base Station ($RBS$), where each agent would use these previous size values to the $k$ internal predictors and calculate the predicted value. If the drone's internal predicted value is less than $B$, the agent decides to recharge the batteries or continue its work.
CT policy also considers the battery values of these drones either at the lower value ($LW$) so that the agent who has a charge lower than this value decides to recharge, or the higher value ($UP$), the battery level value is higher than the UP, the drone will not try to recharge.

\begin{figure}[ht]
	\centering
	\includegraphics[width=0.5\textwidth]{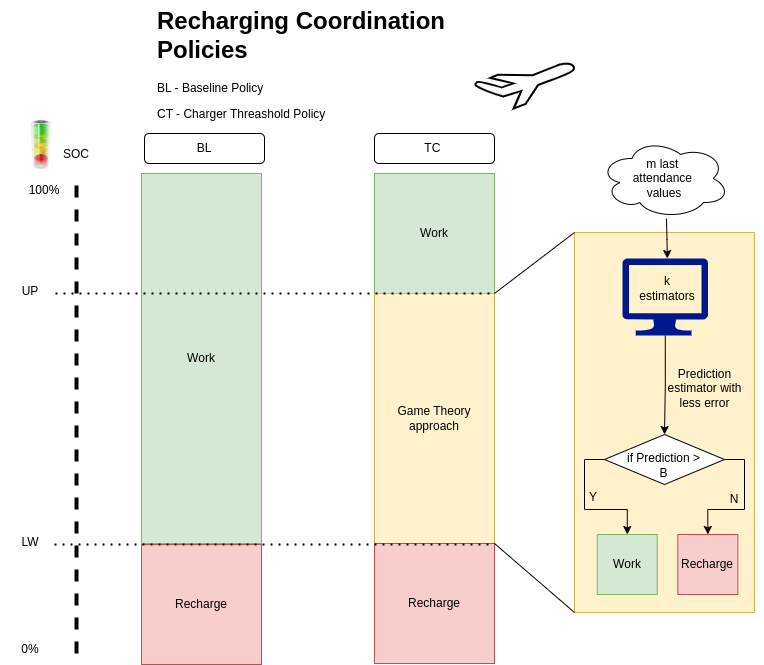} \\
    \caption{Recharge Policy Decision Process.}
	\label{fig:10}
\end{figure}

In the simulation model, the charging station has a limited number of positions. If the value of the attendant drone's number in a given simulation time is more than the $B$ value, this attendant drone will not recharge. 
About the drone battery modeling, the battery State of Charge $(SOC)$ range goes from 0 to 100\% battery level. Two simulation parameters increase or decrease the $SOC$ during the simulation. The first is the Battery Gain $(BG)$ that increases the $SOC$, emulating the battery hot-swapping process in a given time (1 time step). 
In each run, the drones' SOC value decreases in a Battery Consumption $(BC)$ rate. Because drone usage can be randomized by environment and work factors such as UAV payload, wind speed, and UAV vertical and horizontal movement, the model considers a normal standard deviation value $(SD)$ in the Battery Consumption value.
Regarding time dimension, we use Netlogo’s internal clock time step unit, called $“tick”$. Drones move randomly in a 2D space, according to their state defined during the simulation run to recharging or working patches.

\section{Results and discussion}

The experiment consists of a sensitivity analysis of the model depending on battery demand values. The experiment evaluates 120 scenarios (60 scenarios for each policy), with 100 repetitions, resulting in 12000 simulation rounds. Table \ref{tab:simulation_parameters} presents the fixed and variable simulation parameter values. There are two policies ($BL$ and $CT$). The recharging place size $(B)$ has four recharging possibilities: 30\% and 40\% of the $QTY$ value. The $B$ value varies following the $QTY$ value in each $tick$, For example, if the $B$ is set to 20\% and there are 50 remaining drones, the recharging space is 10 drones. The Battery Consumption can take 15 values, going from 1\% to 15\% of $SOC$. This is done to evaluate how policy behavior is better in possible different drone usage conditions.

\begin{table}[ht]
\centering
\caption{Simulations parameters.}
\begin{tabular}{lclcc}
\hline
\textbf{Parameters type} & \textbf{Symbol} & \textbf{Description} & \textbf{Value Range} & \textbf{Levels} \\
\hline
\multirow{8}{*}{Fixed} 
 & $m$   & CT Policy data windows.                      & 10     & 1 \\
 & $k$   & CT policy internal estimator number          & 5      & 1 \\
 & UP    & CT policy upper value                        & 80     & 1 \\
 & LW    & CT and BL policies lower value               & 25     & 1 \\
 & SD    & BC standard deviation                        & 0.1    & 1 \\
 & QTY   & The initial number of drones                 & 100    & 1 \\
 & Ticks & Unity of time                                & 1500   & 1 \\
 & BG    & SOC increase in effective recharging         & 100\%  & 1 \\
\hline
\multirow{3}{*}{Variable}
 & Policy & Drone recharging decision Policies          & BL e CT & 2 \\
 & BC     & SOC decreases in every simulation run        & 1, 2, ..., 15 & 15 \\
 & B      & Ratio of QTY that represents the recharging place capacity & 30, 40 & 2 \\
\hline
\end{tabular}
\label{tab:simulation_parameters}
\end{table}

The stopping criterion for each round is the presence of no agent with a remaining battery or 1500 rounds of simulation.
In agent-based simulation, there are two levels of results to describe the interactions between entities. The micro-level results describe a simple local behavior, and the macro-level, derived from the micro-levels, describes the interaction of more elements \cite{Remondino2006}. 

Three experiments results values were found, concerning the reliability and the efficiency of each simulation set. Reliability results (1 and 2) consider the macro-level results.  The efficiency result (3), in turn, aims at the micro-level result. These results are described as follows:

\begin{enumerate}
    \item \textbf{Average Simulation Remaining Drones} is related to the average number of drones that completed the simulations. It represents that a system will have great mission coverage;
    \item \textbf{Percentage of Finished Simulations} in which all 120 simulation sets completed the 100 repetitions;
    \item \textbf{Average Utility} is the agent work/recharge decision ratio, that is, the ratio of times drones decide to work instead of recharging during all simulations. A higher utility represents that this simulation set does more work than a lower utility.
\end{enumerate}

\subsection{Reliability Compliance}

\begin{figure}[ht]
	\centering
	\includegraphics[width=0.8\textwidth]{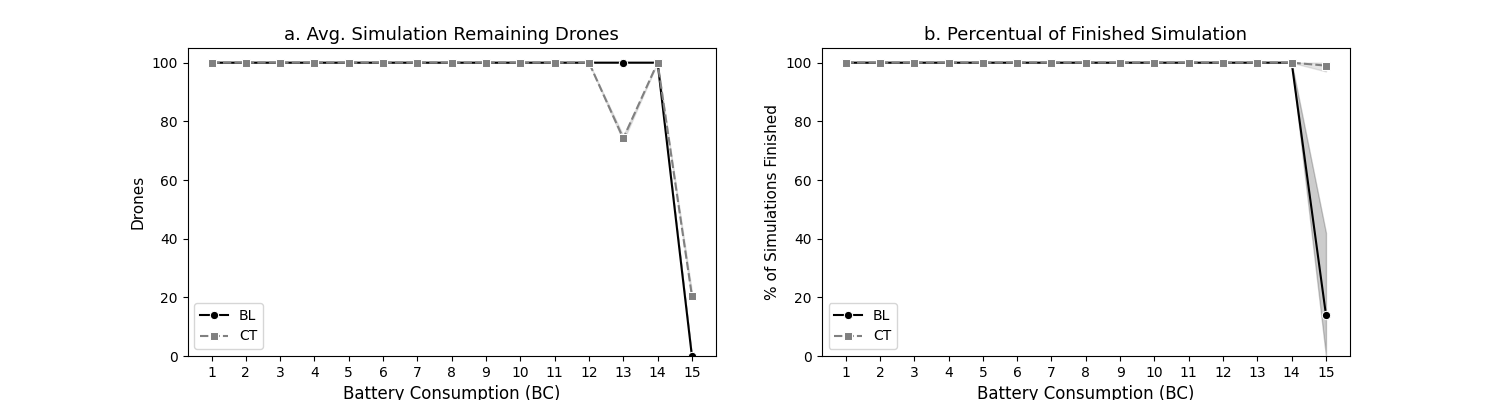} \\
    \caption{Reliability Results Analysis for B = 30\%.}
	\label{fig:11}
\end{figure}

\begin{figure}[ht]
	\centering
	\includegraphics[width=0.8\textwidth]{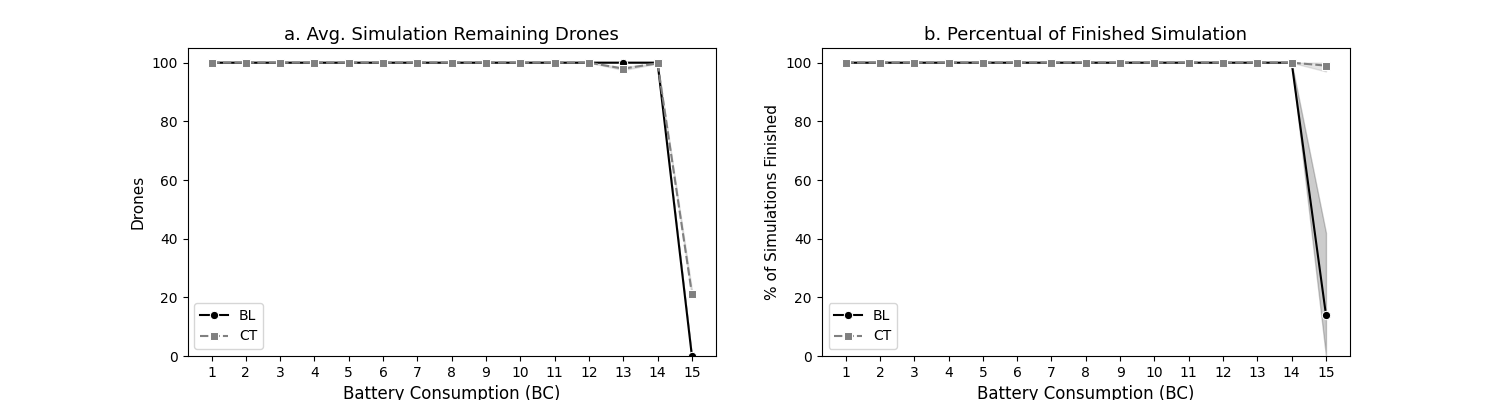} \\
    \caption{Reliability Results Analysis for B = 40\%.}
	\label{fig:12}
\end{figure}

Figures \ref{fig:11}-a, \ref{fig:12}-a  present the Average Simulation of Remaining Drones, and Figures \ref{fig:11}-b, \ref{fig:12}-b present the Percentage of Finished Simulation mean values for each 60 simulation set. 

These results were reliability-related. When the $BC$ values is less than or equal to 12\% of $SOC$ all simulations run results in 100\% of simulations finished, and all drones finish the simulation runs. 
When BC is in extreme condition $(BC > 12)$, simulation setups have a performance degradation, but the baseline policy has a stronger degradation in the case of the hardest condition $(BC = 15\%)$. Because the $LW$ value (25\%) is near the BC and without any other intelligence, the drones with this policy will stop working more than the CT drones policy. About recharging space $(B)$, the effect shows that increasing B size improves the performance of the final result. The CT policy drones are more sensitive about the B change. 

\subsection{Efficiency Analysis}

\begin{figure}[ht]
	\centering
	\includegraphics[width=0.6\textwidth]{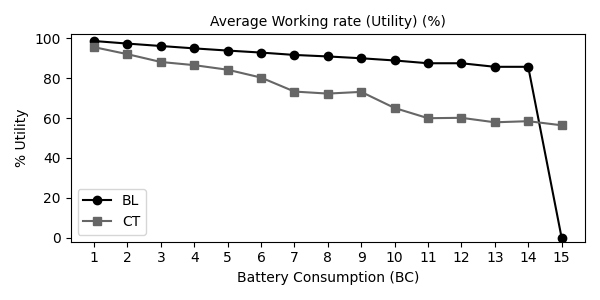} \\
    \caption{Efficiency Analysis Results for B = 30\%.}
	\label{fig:13}
\end{figure}

\begin{figure}[ht]
	\centering
	\includegraphics[width=0.6\textwidth]{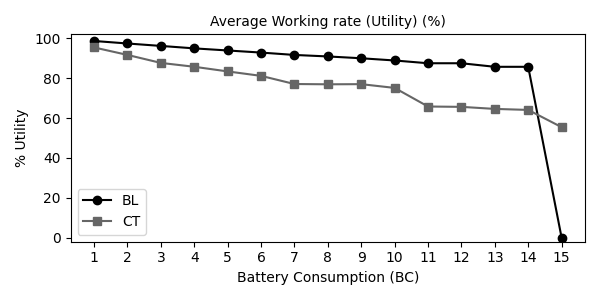} \\
    \caption{Efficiency Analysis Results for B = 40\%.}
	\label{fig:14}
\end{figure}

Figures \ref{fig:13} and \ref{fig:14} present the average utility for each simulation for each $B$ value. The utility is the ratio that the simulation set is working in the 1500 simulation runs. The utility shows a linear decrease in the $BC$ value, That is waited, because more use, more recharging. In the BL police, when $BC$ = 15, the utility is 0 because there are no remaining drones in B = 30, and 40\%. CT policy case, until the utility is less BL, the BC = 15 shows an opportunity for this policy in extreme situations.  

As expected, in the mean simulation runs, the BC increase reduces the simulation set performances, so our model has a valid behavior. The $B$ increase value shows less correlation between simulation sets' performance but more in CT policy. 

\section{Conclusions}

This article aims to describe drone usage in three applications and study a possible study of the energy supply to a swarm of drones, proposing a recharging coordination process using a game theory approach. 
Drones can improve agricultural resources to help farmers make the best decisions. In disaster recovery applications, drones can provide communication, imaging, and evacuation routes to disaster victims. In the dengue combat application, drones can locate mosquito vectors and release infertile males to reduce this disease.
Regarding the recharging procedure, we propose agent-based modeling to evaluate how a swarm of drones behaves in an autonomous scenario regarding their recharging decision. We consider two cases to define the effectiveness of recharging decisions. The first condition is the Baseline (BL) Policy, in which drones make decisions about their battery level, and the Charger Threshold (CT) Policy uses the battery level and a game theory approach called El Farol Bar. We made an abstraction in the development of a simulation workplace, battery usage with a standard deviation to emulate several drone uses, and energy supply and evaluated two types of results, reliability and efficiency results. 
Regarding reliability system compliance in the case of low UAV usage, there are no differences between policies. But in the $BC = 15\%$ case, CT policy best performs. In the efficiency results case, policies had good results, with agents working in more than 80\% of the simulation. But as reliability results in the case of more extreme situations, TC policy agents had better performance.
This work shows the opportunity to deploy a swarm to work applications and others. The workforce required to manipulate and control these devices is expensive, and internal intelligence proposed for our policies can help in the future automation of these devices. 

\subsection{Future Works}

For future works, we propose the addition of parameters related to working or recharging decisions. These parameters will deliver realism in the validation process. Other agents' internal process decisions can be simulated. The recharging process can be improved by adding a queue for the recharging place. Future works may also consider other recharging modes, such as drone swapping. Finally, in future models, it may be interesting to consider the network properties for drone-drone and drone-ground communications. 

\section*{Acknowledgments}

This study was financed in part by the Coordenação de Aperfeiçoamento de Pessoal de Nível Superior - Brasil (CAPES) - Finance Code 001.

\bibliographystyle{unsrt}  
\bibliography{template}  



\end{document}